\documentclass[12pt]{article}

\begin{document}

\title{Hydrodynamic time correlation functions in the presence of a
gravitational field}
\author{A. Sandoval-Villalbazo$^a$ and L.S. Garc\'{\i}a-Col\'{\i}n$^{b,\,c}$ \\
$^a$ Departamento de Ciencias, Universidad Iberoamericana \\
Lomas de Santa Fe 01210 M\'{e}xico D.F., M\'{e}xico \\
E-Mail: alfredo.sandoval@uia.mx \\
$^b$ Departamento de F\'{\i}sica, Universidad Aut\'{o}noma Metropolitana \\
M\'{e}xico D.F., 09340 M\'{e}xico \\
$^c$ El Colegio Nacional, Centro Hist\'{o}rico 06020 \\
M\'{e}xico D.F., M\'{e}xico \\
E-Mail: lgcs@xanum.uam.mx}
\maketitle

\begin{abstract}
{\small This paper shows that the ordinary Brillouin spectrum peaks
associated to scattered radiation off acoustic modes in a fluid suffer a
shift in their values due to gravitational effects. The approach is based in
the ordinary linearized Navier-Stokes equations for a fluid coupled to a
Newtonian gravitational potential. The formalism leads to a dispersion
relation that contains both gravitational and dissipative effects. It is
also shown that the Brillouin peaks tend to condense into a single peak when
the fluid modes approach the critical Jeans wave number.}
\end{abstract}

\section{\textbf{Introduction}}

Dynamic light scattering theory ~\cite{Mountain}-\cite{Berne}
shows that photons interact with the acoustic modes in a fluid.
This scattering arises from sound waves associated to statistical
fluctuations in the fluid's density and temperature. Early
theoretical work in this subject, as well as the measurements
confirming the existence of the Brillouin-Rayleigh (BR) spectra
are given in Refs. [1]-[3]. The first theoretical approach to the
subject is a lapidal note published by Landau and Placzek back in
1936 [4] which has been the inspiration of most of the work done
on this problem. Recently, these ideas have been extended into the
realm of astrophysical and cosmological contexts in relation with
the discussion of the old problems on gravitational instabilities
and Jeans' mass number [5] \cite{Groot1}. The recognized fact that
Cosmic Microwave Background Radiation (CMBR) photons interact with
acoustic fluid modes in certain stages of the universe evolution
suggests that Brillouin scattering is potentially important to
CMBR physics.

On the other hand, entropy production plays an important role in
realistic descriptions of thermodynamical processes, and some
discussion of its physical sources may yield interesting results
in the analysis of the time evolution of statistical fluctuations
in astrophysical systems \cite{Weinberg}. Recent work
~\cite{Mimismo} shows indeed that the critical Jeans wave number
for gravitational collapse is slightly modified by dissipative
effects such as viscosity. The formalism used in that work is also
useful to obtain structure factors, namely \emph{scattering laws}
that describe line broadenings and frequency shifts associated to
the interaction between the incoming radiation and the fluid
modes. The mathematical tool in this case is the so-called
density-density correlation function, taken in the frequency
(Fourier) domain which is obtained by standard techniques~\cite
{Berne}-[3]. The influence of the gravitational field in the
Brillouin spectrum will become apparent from the analytic form of
the structure factor.

In this paper we wish to discuss two results. Firstly, that the presence of
a gravitational field produces a shift in the effective frequencies of the
Brillouin doublet in a simple fluid. The shift could be significant for
large enough densities such as those of matter in early stages of the
Universe. Secondly, the Brillouin peaks tend to condense into a single peak
as the critical Jeans wave number is approached. This implies that at
certain densities, the thermal fluctuations and the ``mechanical dissipative
modes'' behave more or less in the same way, a fact which, if detectable,
should become manifest in the anisotropies of the observed CMBR. The
significance of this fact has not yet been explored.

This paper is thus divided as follows. Section two is dedicated to
a review of the linearized form the Navier-Stokes equations in the
presence of a gravitational field. Section three is devoted to the
derivation of the scattering law and its comparison with the
ordinary BR spectrum. A brief discussion of the implications of
the results here obtained is included in section four.

\section{Linearized Navier-Stokes system in the presence of a gravitational
field}

The standard system describing fluctuations of the
thermohydrodynamical variables in a simple non-reacting fluid
consisting of particles of mass $m$ in the presence of a
gravitational field, is derived by following the tenets of Linear
Irreversible Thermodynamics (LIT)~\cite{Berne}~\cite{Mimismo}.
Indeed,the so-called Navier-Stokes-Fourier equations of
hydrodynamics for a simple fluid arise from the structure of what
is now called LIT by supplementing the two conservation equations
for mass and momentum, respectively and the balance equation for
the internal energy, with additional information. Indeed, these
equations read ~\cite{Berne}~\cite{Groot1}-\cite{GC}.

\begin{equation}
\frac{\partial \rho }{\partial t}+\frac{\partial }{\partial
x^{i}}(\rho \,u^{i})=0  \label{Nav1}
\end{equation}
\begin{equation}
\rho \frac{D\,\,u^{i}}{Dt}+\frac{\partial \Xi ^{\,i\,j}}{\partial
x^{j}} =F^{i}  \label{Nav2}
\end{equation}
\begin{equation}
\rho \frac{D\,\,\varepsilon }{Dt}+\frac{\partial Q^{\,j}}{\partial
x^{j}} =-\Xi ^{\,i\,j}\,\frac{\partial u_{i}}{\partial x^{j}}
\label{Nav3}
\end{equation}
where $\frac{D\,}{Dt}=\frac{\partial }{\partial
\,t}+u^{i}\frac{\partial }{\partial x^{i}}.$

\bigskip Here, $\rho (x^{j},t),\,u^{i}(x^{j},t)$ and $\varepsilon (x^{j},t)$
are the local density, velocity and internal energy, respectively,
$\Xi ^{\,i\,j}$ the momentum current (or stress tensor) and
$Q^{\,j}$ the heat flux. All indices run from 1 to 3. In general,
for isotropic fluids, $\Xi ^{\,i\,j}=p\delta ^{\,i\,j}+\tau
^{\,i\,j}$, where $p$ is the local hydrostatic pressure and $\tau
^{\,i\,j}$ the viscous tensor ($\delta ^{\,i\,j}$ is the unitary
dyadic). Notice that Eqs. (\ref{Nav1}-\ref{Nav3}) contain fifteen
unknowns (including the pressure) and there are only five
equations, so the system is not well determined. If we arbitrarily
choose to describe the states of the fluid through the set of
variables $\rho (x^{j},t),\,u^{i}(x^{j},t),$ $T(x^{j},t),$ where
$T$ is the local temperature, we need nine dynamic equations of
state (or constitutive equations) relating $\Xi ^{\,i\,j}$ and
$Q^{\,j}$ to the state variables plus two local equations of state
$p=p(\rho ,T)$ and $\varepsilon =\varepsilon (\rho ,T).$ According
to the tenets of LIT we choose the so-called linear constitutive
laws, namely
\begin{equation}
\tau ^{\,i\,j}=-2\eta \,\sigma ^{\,i\,j}-\zeta \theta \delta
^{\,i\,j} \label{DSE1}
\end{equation}
\begin{equation}
Q^{\,j}=-\kappa \delta ^{\,ji}\frac{\partial T}{\partial x^{i}}
\label{DSE3}
\end{equation}
with $\eta $ and $\zeta $, the shear and bulk viscosities,
respectively, $ \sigma ^{\,i\,j}$ the symmetrical traceless part
of the velocity gradient,  $ \theta =\frac{\partial
u^{i}}{\partial x^{i}}$ and $\kappa $ being the thermal
conductivity.

Eqs. (\ref{DSE1}) and (\ref{DSE3}) are the well known constitutive
equations of Navier-Stokes and Fourier, respectively. Substitution
of these equations into Eqs. (\ref{Nav1}-\ref{Nav3}) yields a set
of second order in space, first order in time non-linear coupled
set of partial differential equations for the chosen variables
$\rho ,\,u^{i}$ and $T$. This set, which the reader may seek in
the literature \cite{Berne},\cite{Groot1}-\cite{GC}, is the so
called Navier-Stokes-Fourier system of hydrodynamic equations.The
non-linearities appearing in such equations have two sources, the
inertial terms $\,u^{i} \frac{\partial }{\partial x^{i}}$ \
arising from the hydrodynamic time derivatives, plus quadratic
terms in the gradients of velocity arising from Eqs. (\ref{DSE1})
and (\ref{DSE3}). Moreover, it should be mentioned that this set
of equations is consistent with the second law of thermodynamics,
the Clausius uncompensated heat , or entropy production, is
strictly positive definite.

\bigskip

Nevertheless, for the purpose of this paper, this set of equations
is too complicated. In order to deal with fluctuations around the
equilibrium state, one assures that for any of two state
variables, call them $X(x^{j},t) $ one can write that,
\begin{equation}
X(x^{j},t)=X_{o}+\delta X(x^{j},t)  \label{fluctuations1}
\end{equation}
where $X_{o}$ is the equilibrium value of $X$ and $\delta X$ the
corresponding fluctuation. Neglecting all terms of order $(\delta
X(x^{j},t))^{2}$ and higher in the NSF non-linear set one finds
the {\em linearized NSF equations of hydrodynamics}
\cite{Berne}-\cite{GC},
\begin{equation}
\frac{\partial }{\partial t}\left( \delta \rho \right) +\rho
\,\theta =0 \label{Fluc1}
\end{equation}
\begin{equation}
\rho _{o}\frac{\partial u_{k}}{\partial t}=-\,\frac{1}{\rho
_{o}\,\kappa _{T} }\frac{\partial }{\partial x^{k}}\left( \delta
\rho \right) -\frac{\beta }{ \kappa _{T}}\frac{\partial }{\partial
x^{k}}\left( \delta T\right) +2\eta \nabla^{2}u_{k}-(\frac{2}{3
}\eta -\varsigma )\frac{\partial }{\partial x^{k}}(\theta )+F_{k}
\label{Fluc2}
\end{equation}
\begin{equation}
\frac{\partial }{\partial t}\left( \delta T\right) =D_{T}
\nabla^{2}\delta T-\frac{ \beta \,T}{\rho _{o}C_{p}\kappa
_{T}}\theta \label{Fluc3}
\end{equation}
since $u_{k\;o}=0$, $u_{k}=\delta u_{k}$ and $\frac{\beta }{\kappa
_{T}}=( \frac{\partial p_{0}}{\partial T_{o}})_{\rho _{o}}$,
$\kappa _{T}=\rho _{o}( \frac{\partial p_{0}}{\partial \rho
_{o}})_{T_{o}}$ and $D_{T}=\frac{k}{\rho _{o}C_{v}}$ is the
thermal diffusivity. $C_{p}$ and $C_{v}$ are the specific heats at
constant pressure and constant volume, respectively.

Note now that from Eq. (\ref{Fluc2}), $u_{k\;Trans}\equiv
Rot\left( u_{k}\right) $ uncouples from the hydrodynamic modes,
whence, one formally arrives at the desired set, namely:
\begin{equation}
\frac{\partial }{\partial t}\left( \delta \rho \right) +\rho
_{o}\,\theta =0
\end{equation}
\begin{equation}
\rho _{o}\frac{\partial \theta }{\partial t}=-\,\frac{1}{\rho
_{o}\,\kappa _{T}} \nabla^{2}\delta \rho-\frac{\beta }{\kappa
_{T}} \nabla^{2}\delta T+D_{v}\nabla^{2}\delta
\theta-\nabla^{2}\delta \varphi \label{cinco}
\end{equation}
\begin{equation}
\frac{\partial }{\partial t}\left( \delta T\right)
=D_{T}\nabla^{2}\delta T-\frac{ \beta \,T_{o}}{\rho
_{o}C_{p}\kappa _{T}}\theta   \label{seis}
\end{equation}
assuming that $F_{k}=-\frac{\partial \,\varphi }{\partial x^{k}}$,
where $\varphi$ is the gravitational potential. Also, $
D_{v}=(\frac{4}{3}\eta +\varsigma )\frac{1}{\rho _{o}}$. Now, we
notice that $\frac{\partial }{\partial t}\left( \delta \rho
\right) =-\rho _{o}\,\theta $ so we can reduce this set to only
two equations, Eq.(\ref{cinco}) and Eq.(12).  We now perform a few
minor transformations introducing the speed of sound of the fluid
$C_{o}^{2}$ $=\left( \frac{\partial p}{\partial \rho }\right)
_{s}$ through the relationship
\begin{equation}
\kappa_{T}=\gamma \kappa_{s}=\frac{C_{p}}{C_{v}}\kappa_{s}
\label{Sound1}
\end{equation}
where $k_{s}$ is the adiabatic compressibility $\left(
\frac{\partial \rho }{
\partial p}\right) _{s}$ whence, $k_{T}=\frac{\gamma }{C_{o}^{2}}$. This
leaves us finally with the set:
\begin{equation}
-\frac{\partial ^{2}(\delta \rho )}{\partial
t^{2}}+\frac{C_{o}^{2}}{\gamma } \nabla^{2}\delta
\rho+\frac{C_{o}^{2}\beta }{\gamma } \nabla^{2}\delta
T+D_{v}\nabla^{2}(\frac{\partial }{
\partial t}\left( \delta \rho \right) )+\rho _{o}\nabla^{2}\delta \varphi=0  \label{sis1}
\end{equation}
\begin{equation}
\frac{\partial }{\partial t}\left( \delta T\right)
-D_{T}\nabla^{2}\delta T-\frac{ \gamma -1}{\beta }\frac{\partial
}{\partial t}\left( \delta \rho \right) =0 \label{sis2}
\end{equation}
where we have used that $C_{p}-C_{v}=\frac{\beta ^{2}\,T_{o}}{\rho
_{o}^{2}\kappa _{T}}$.

\bigskip Eqs.(\ref{sis1}-\ref{sis2}) form a set of coupled equations for the
density and temperature fluctuations in the fluid under the action
of a conservative force whose nature need not to be specified for
the time being. They are the basis for studying the properties of
the time correlation functions of thermodynamic fluctuations.
Those of the density will be of particular interest here.

\section{Solution to the hydrodynamic equations}

The results derived in the previous section are far from being
new. Aside from the term $\nabla ^{2}\phi $ which arises from the
presence of an external conservative force, they are identical to
the ones that have been widely discussed in the literature. The
question here is if the gravitational potential introduces any
substantial modification in the correlation functions for the
thermodynamic fluctuations. To examine this possibility we recall
that, if we are considering only the fluctuations, the
gravitational potential satisfies the Poisson equation:

\begin{equation}
\nabla^{2}\delta \varphi=-4\pi G\,\delta \rho   \label{Sol1}
\end{equation}

Now, equation (\ref{sis2}) reads
\begin{equation}
-\frac{\partial ^{2}(\delta \rho )}{\partial
t^{2}}+\frac{C_{o}^{2}}{\gamma } \nabla^{2}\delta
\rho+\frac{C_{o}^{2}\beta } \gamma \nabla^{2}\delta T
+D_{v}\nabla^{2} (\frac{
\partial }{\partial t}\left( \delta \rho \right) )-4\pi G\rho _{o}\,\delta
\rho =0  \label{sis3}
\end{equation}
The introduction of the Poisson equation links fluctuations in the
gravitational potential with density fluctuations. The solution to
Eqs.(\ref{sis2}) and (\ref{sis3}) proceeds in the standard
fashion. We reduce them to a set of algebraic equations by taking
their Laplace-Fourier transform, choose to set the static
temperature fluctuations equal to zero, and eliminate the
temperature leading to an equation for $\delta \rho \left(
\vec{k},s\right) $, which is the ratio of two polynomials in $s$.
In fact, one gets that

\begin{equation}
\frac{\delta \rho (\vec{k},s)}{\delta \rho (\vec{k},0)}=\frac{
s^{2}+(D_{v}+D_{t})k^{2}s+k^{4}D_{v}D_{t}+C_{o}k^{2}(1-\frac{1}{\gamma
})}{ (s+D_{t}\,k^{2})(s^{2}+D_{v}k^{2}s+C_{s}^{2}k^{2}-4\pi G\rho
_{o})} \label{dispnew}
\end{equation}

To compute $\delta \rho \left( \vec{k},t\right) $ one must take
the inverse Laplace transform of the former quantity, which
demands the knowledge of the roots of the denominator, which is a
cubic equation in $ s $ (dispersion equation).

The analysis leading to Eq.(\ref{dispnew}) clearly points out the
fact that the solution to the cubic equation giving rise to the
poles of the function $\delta \rho (\vec{k},s)$, is exact. This is
an improvement over the current version in the literature
\cite{Berne}-\cite{Boon}, \cite{GC} asserting that the roots are
only approximate to order $k^{2}$. There are two immediate
implications of this result, namely, Rayleigh's peak is not
affected by the gravitational field within the Navier-Stokes
regime and the Jeans number, thoroughly discussed in
Ref.\cite{Mimismo} cannot be affected by thermal dissipation, the
thermal conductivity will never appear in its definition, as will
be pointed out below. The former result appears to be in
contradiction with previous work stating that Rayleigh's peak is
modified by a constant gravitational force \cite{Sengers}. Yet,
notice that this is a $k^{4}$ effect, which is beyond the
Navier-Stokes domain. Work along this line with the Burnett and
SuperBurnett equations is in progress. So, we shall leave the
discussion of this subtlety for the future.

Returning to Eq. (\ref{dispnew}), the two roots of the quadratic
equation in the denominator are:
\begin{equation}
s_{1,2}=-\frac{D_{v}k^{2}}{2}\pm i\left[ \left(
C_{o}^{2}k^{2}-4\pi G\rho _{o}\right)
-\frac{D_{v}^{2}k^{4}}{4}\right] ^{1/2}  \label{rootsnew}
\end{equation}
If $4\pi G\rho _{o}=0$ and viscosity dominates over the term in
$k^{2}$, this result reduces to the one giving rise to the
standard Brillouin peaks which correspond to density fluctuations
of the type \cite{Berne} \cite{GC},
\begin{equation}
\delta \rho \left( \vec{k},t\right) =\delta \rho \left(
\vec{k},0\right) \frac{1}{\gamma }e^{-D_{v}k^{2}t}Cos\left[
C_{o}k\,t\right] \label{Brillouin}
\end{equation}
which are the acoustic modes damped by the Stokes-Kirchhoff factor
$D_{v}$ .

On the other hand, if $4\pi G\rho _{o}\neq 0$, the threshold value
for $k$ distinguishing between damped oscillations and growing
modes is given by
\begin{equation}
\left( C_{o}^{2}k^{2}-4\pi G\rho _{o}\right)
-\frac{D_{v}^{2}k^{4}}{4}=0 \label{Tnew}
\end{equation}
or,
\begin{equation}
k^{2}=\frac{2C_{o}^{2\,}}{D_{v}^{2}}\left( 1\pm \sqrt{1-\frac{4\pi
G\rho _{o}D_{v}^{2}}{C_{o}^{4\,}}}\right)   \label{threshold2}
\end{equation}
Eq. (\ref{threshold2}) is a generalization of Jeans wave number
when dissipative effects due to viscosity are non-negligible, and
is the main result of the paper. We can note that, if $\frac{4\pi
G\rho _{o}D_{v}^{2}}{ C_{o}^{4\,}}\ll 1,$ and taking the $-$ sign
for the square root, we have:
\begin{equation}
k^{2}\approx \frac{2C_{o}^{2\,}}{D_{v}^{2}}\left[
1-1+\frac{1}{2}\left( \frac{4\pi G\rho
_{o}D_{v}^{2}}{C_{o}^{4\,}}\right) \right]   \label{limit1}
\end{equation}
or
\begin{equation}
k^{2}\approx \frac{4\pi G\rho _{o}}{C_{o}^{2\,}}=K_{J}^{2}
\label{limit2}
\end{equation}
This is the square value of the Jeans wave number. It was derived
by Jeans in 1902 and rederived in many other waves by several
authors. Here we simply show that it is almost a trivial
consequence of LIT. Jeans wave number has been used by
cosmologists to estimate the minimum mass required to form a
galaxy. This question has been discussed at length in Ref.
\cite{Kolb}.

Coming back to our main result, Eq. (\ref{dispnew}), it is clear
that damped acoustic waves are associated to the case in which the
dispersion equation has complex roots. The corresponding
expression for the density fluctuations in the $(\mathbf{k},t)$
space, neglecting  the last term in the square root of Eq.
(\ref{rootsnew}), is then given by:

\begin{equation}
\delta \rho (\mathbf{k},t)=\delta \rho (\mathbf{k},0)\left\{ \left( 1-\frac{1
}{\gamma }\right) e^{-D_{T}\,k^{2}t}+\frac{1}{\gamma }e^{-\Gamma \,k^{2}t}Cos
\left[ \left( C_{o}^{2}k^{2}-4\pi G\rho _{o}\right) ^{1/2}\,t\right] \right\}
\label{seis}
\end{equation}

Eq. (\ref{seis}) represents damped acoustic waves propagation with an
effective frequency $\left( C_{o}^{2}k^{2}-4\pi G\rho _{o}\right) ^{1/2}$.
When gravity is negligible, one recovers the usual frequency $\omega =\pm
C_{o}k$. Notice, however, that although $G$ is a small number, it is
multiplied by the density $\rho _{o}$. In fact, it may turn that for a
critical value of $\rho _{o}$ the damped modes are replaced by growing modes
(Jeans instability).

\section{Structure factor and modified Brillouin spectrum}

Following the standard dynamic radiation scattering theory
\cite{Berne}-[3], one may then construct the density-density
correlation function from Eq.(\ref{seis}). This leads to the
expression:

\begin{equation}
\frac{\left\langle \delta \rho ^{\ast }(\mathbf{k},0)\delta \rho
(\mathbf{k} ,t)\right\rangle }{\left\langle \delta \rho ^{\ast
}(\mathbf{k},0)\delta \rho (\mathbf{k},0)\right\rangle }=\left(
1-\frac{1}{\gamma }\right) e^{-D_{T}\,k^{2}t}+\frac{1}{\gamma
}e^{-\Gamma \,k^{2}t}Cos\left[ \left( C_{o}^{2}k^{2}-4\pi G\rho
_{o}\right) ^{1/2}t\right]  \label{siete}
\end{equation}
where the brackets $<\,>$ indicate an average over an equilibrium ensemble.

The Brillouin specific intensity of the scattered light due to its
interaction with the acoustic modes of a fluid is obtained by Fourier's
transform of Eq. (\ref{siete}). The ensuing expression reads:

\begin{equation}
\begin{array}{c}
S_{\rho \rho }\left( \mathbf{k},\omega \right) =\frac{1}{\pi }[(1-\frac{1}{
\gamma })\frac{D_{T}\,k^{2}}{\omega ^{2}-D_{T}^{2}\,k^{4}}+\frac{1}{\,\gamma
}(\frac{\Gamma \,k^{2}}{\left( \omega -\left( C_{o}^{2}k^{2}-4\pi G\rho
_{o}\right) ^{1/2}\right) ^{2}+\Gamma ^{2}\,k^{4}} \\
+\frac{\Gamma \,k^{2}}{\left( \omega +\left( C_{o}^{2}k^{2}-4\pi G\rho
_{o}\right) ^{1/2}\right) ^{2}+\Gamma ^{2}\,k^{4}})]
\end{array}
\label{nueveb}
\end{equation}

The last two terms in Eq. (\ref{nueveb}) correspond to the
Brillouin doublet and deserve some attention. This doublet
reflects a shift in frequency governed by the speed of sound of
the fluid, the wave number of the incoming radiation, and the
effect of gravity. This last effect can be roughly estimated by
performing a binomial expansion of the square root included in the
effective wave frequency. Indeed, let $\omega _{p}$ , the modified
frequency be defined as:

\begin{equation}
\omega _{p}=\pm \left( C_{o}^{2}k^{2}-4\pi G\rho _{o}\right)
^{1/2}\simeq C_{o}k\left( 1\mp \frac{2\pi G\rho _{o}}{C_{o}^{2}k^{2}}
\right)  \label{nueve}
\end{equation}
to first order in $G\rho _{o}$. Thus, if we now call $\left| \Delta \nu
_{p}\right| =\frac{1}{2\pi }\left| \omega -\omega _{p}\right| $ we see that

\begin{equation}
\left| \Delta \nu _{p}\right| =\frac{G\rho _{o}}{C_{o}k}  \label{diez}
\end{equation}
and the relative change in the frequency Brillouin peaks is, roughly:

\begin{equation}
\frac{\left| \Delta \nu _{p}\right| }{C_{o}k}=\left( \frac{G\rho _{o\;}c^{2}
}{4\pi ^{2}C_{o}^{2}}\right) \frac{1}{\nu }  \label{doce}
\end{equation}
if using $k=\frac{2\pi }{\lambda }$ $=\frac{2\pi }{c}\nu $. Written in this
way the frequency exhibits a much subtler dependency on other variables such
as the temperature, which implicitly appears in $C_{o}$.

\section{Final remarks}

A rather interesting result appears in Eq. (\ref{nueveb}) if $k=k_{J}=\frac{
4\pi G\rho _{o\;}}{C_{o}^{2}}$, the Jeans wave number. In that case, $\delta
\rho (\mathbf{k},t)$ is a simple decaying exponential with time, and the
Brillouin-Rayleigh spectra reduces to the ordinary Rayleigh peak superposed
to a similar one arising from the collapse of the Brillouin peaks,
suggesting that the Stokes-Kirchhoff dissipative factor behaves in a way
similar to the thermal fluctuations. This collapse of the Brillouin peaks
would cause a line broadening in scattering situations and would reflect a
competition between entropy fluctuations $(D_{T})$ and mechanical
fluctuations $(D_{v}).\,$Whether or not this result is of any relevance at
all remains to be tested.

A final remark concerns Eq.(\ref{diez}). It is well-known that for typical
electromagnetic radiation, the frequencies are rather high, even radio
frequencies. This is not the case for gravitational waves, whose wavelengths
are very long. One could then speculate about the possibility of having
gravitational radiation Brillouin scattered by cosmological matter. This
would enhance the relative change in frequency of the Brillouin peaks.

This work has been supported by CONACyT (Mexico), project 41081-F.


\begin{thebibliography}{1}
\bibitem[1]{Mountain} R.M. Mountain, Rev. Mod. Phys. \textbf{38}, 205 (1966);

\bibitem[2]{Berne} B.J. Berne and R. Pecora, \textit{Dynamic Light Scattering}
(Dover, N.Y., 2000);

\bibitem[3]{Boon} J.P. Boon and S. Yip, \emph{Molecular Hydrodynamics}
(Dover, N.Y., 1991).

\bibitem[4]{Landau1} L. Landau and G. Placzek, Phys. Zeit. Sow. \textbf{5},
172 (1934). Also included in, \textit{The Collected Papers of L.D. Landau, }
ed. D. ter Haar (Pergamon Press, London, 1962).

\bibitem[5]{lc} A. Sandoval-Villalbazo and R. Maartens
(2001)[astro-ph/0105323]

\bibitem[6]{Groot1}  S.R. de Groot and P. Mazur,(1984) {\it Non-Equilibrium
Thermodynamics}, Dover, N.Y., USA.

\bibitem[7]{GC}  L.S. Garc\'{i}a-Col\'{i}n, {\it Termodin\'{a}mica de los
Procesos Irreversibles} (UAM-Iztapalapa, M\'{e}xico D.F.1990) (in
spanish).

\bibitem[8]{Sengers}  P.M. Sergh, R. Schmitz and J.V. Sengers,
Physica A \textbf{195}, 31 (1993).

\bibitem[9]{Kolb} E.W. Kolb and M.S. Turner , (1990) {\it The Early
Universe} Addison-Wesley, Reading, MA., USA.

\bibitem[10]{Mimismo} A. Sandoval-Villalbazo and L.S. Garc\'{\i}a-Col\'{\i}n,
Class. Quantum. Grav. \textbf{19} (2002) 2171 [astro-ph/0110295]

\bibitem[11]{Weinberg} S. Weinberg. Ap. J. \textbf{168}, 175 (1971).

\bibitem[12]{nueve} A. Sandoval-Villalbazo; Physica A \textbf{313}, 456
(2002).


\end{thebibliography}
\end{document}